# Channel Estimation and Linear Precoding in Multiuser Multiple-Antenna TDD Systems

Jubin Jose, *Student Member, IEEE,* Alexei Ashikhmin, *Senior Member, IEEE,*
Phil Whiting, and Sriram Vishwanath, *Senior Member, IEEE*

*Abstract*—Traditional approaches in the analysis of downlink systems decouple the precoding and the channel estimation problems. However, in cellular systems with mobile users, these two problems are in fact tightly coupled. In this paper, this coupling is explicitly studied by accounting for channel training overhead and estimation error while determining the overall system throughput. The paper studies the problem of utilizing imperfect channel estimates for efficient linear precoding and user selection. It presents precoding methods that take into account the degree of channel estimation error. Information-theoretic lower and upper bounds are derived to evaluate the performance of these precoding methods. In typical scenarios, these bounds are close.

*Index Terms*—Cellular downlink, channel estimation, linear precoding, wireless communication

## I. INTRODUCTION

There is a rich and varied literature in the domain of multiple antenna cellular systems. Ever since the introduction of multi-antenna systems, almost every combination of antennas with physical settings has been modeled and analyzed. The bulk of this literature, however, has focused on developing strategies for frequency division duplex (FDD) systems, and not without good reason. FDD systems have dominated deployment, while interest in deploying time division duplex (TDD) systems has grown only in recent years. Although TDD and FDD seem like interchangeable architectural schemes for cellular systems, there are some fundamental differences that need to be isolated and studied in detail. The goal of this paper is to bring the understanding of TDD systems closer to that of FDD systems today.

It is now well established that multiple antennas at the transmitter and receiver in a point-to-point communication system can greatly improve the overall throughput of the system [2], [3]. In a multi-user setting, this gain requires channel state information (CSI) and precoding strategies that use this CSI at the basestation. Given this CSI, the channel capacity problem can be formulated in terms of a multi-antenna Gaussian broadcast channel (BC). Over the past decade, the capacity of a multi-antenna Gaussian BC has been determined, and shown to be achieved by using dirty paper coding (DPC) in [4], [5], [6], [7], [8]. Subsequently, the order growth in the sum capacity gain with the number of antennas and the signal to noise ratio (SNR) have been characterized in [9], [10]. An overview of the capacity results in multi-user multiple-input multiple-output (MIMO) channels can be found in [11].

Although dirty paper coding is known to be capacity achieving with perfect CSI, there are several issues when attempting to apply it directly to a cellular system. First, practical systems have to cope with rapidly changing channels so that channel estimates are valid only for a very short time, making the application of DPC a fraught problem. Furthermore, we are mainly concerned with systems that have a large number of base-station antennas. In such systems, the use of DPC might turn out to be prohibitively complex. In contrast, many antenna systems with linear precoding offers a much more practical route to provide high rate wireless communications. Estimation error is an inevitable issue for the linear precoded system (as well as for DPC) and so the paper concentrates on this question. Detailed investigation of DPC performance with channel estimates obtained from TDD pilots remains a question for further research.

Given that we use linear precoding, the goal of this paper is to analyze a multi-antenna downlink TDD system with channel training and estimation error factored into the net throughput expression. One of the primary differences between TDD and FDD systems is the means through which channel training and resulting estimation is conducted. In FDD systems, a common means of gaining CSI is feedback from the users to the basestation. In TDD systems, *channel reciprocity* can be used to train on reverse link and obtain an estimate of the channel at the base-station, see for example [12], [13]. In [14], channel reciprocity has been validated through experiments. Reciprocity thus eliminates the need for a feedback mechanism (along with forward training) to be developed. In literature, the study of joint precoding and feedback schemes for FDD systems have been studied in great detail [15], [16], [17], [18], [19] (see prior work section for details). In a similar vein, we find that a joint study of channel estimation and precoding for TDD systems is needed to understand the resulting overall system throughput. To provide some typical system parameters, consider a carrier frequency of 1900 MHz and (maximum) mobile velocity of 150 miles/hour. Then, the coherence time is approximately $400$ $\mu$s [20]. With typical coherence bandwidth of $50 - 200$ kHz, the effective symbol

Results in this paper were presented in part at the IEEE International Conference on Communications (ICC) 2008 [1].

J. Jose and S. Vishwanath are with the Department of Electrical and Computer Engineering, The University of Texas at Austin, Austin, TX 78712 USA (email: jubin@austin.utexas.edu; sriram@austin.utexas.edu). We thank support from the National Science Foundation under contract CNS-0905200 and the department of defense.

A. Ashikhmin and P. Whiting are with Bell Laboratories, Alcatel-Lucent Inc., Murray Hill, NJ 07974 USA (email: aea@research.bell-labs.com; pwhiting@research.bell-labs.com).

rates for narrow-band operation is approximately $5-20$ $\mu$s. This leads to short coherence time in symbols of $20-80$ symbols, which clearly motivate our joint study of channel training, channel estimation and precoding.

Our analytical framework considers a downlink system with a large number of base-station antennas (along the lines of the framework studied in [21]). In this framework, our focus is not on systems specified by current standards such as WiMax and LTE that use only $2-4$ antennas. Instead, our focus is on possible future generations of wireless systems where an antenna array with a hundred or more antennas at the base-stations is an attractive approach. Preliminary feasibility studies show that for $120$ antennas we need a space occupied by a cylinder of one meter diameter and one meter high: half-wavelength circumferential spacing of $40$ antennas in each of three rings, each ring spaced vertically two wavelengths apart. With such systems, TDD offers a significant advantage over FDD operation. In FDD systems, the forward training overhead needed increases with the number of base-station antennas. This overhead also increases the (limited) feedback needed to gain CSI at the basestation which is often neglected when FDD systems are analyzed. In contrast to this, in this paper, we account for all channel training overhead incurred in the throughput analysis we present.

The main contributions of this paper are:

- We determine a method of linear precoding and user selection that maximize net throughput for realistic TDD systems. That is, channel estimation and the consequent errors are taken into account.
- Our results allow us to optimize the training period in such TDD systems. In other words, we determine the optimal trade-off between estimating the channel and using the channel.
- We provide achievable schemes and upper bounds on the system throughput for the suggested precoding and user selection schemes. We demonstrate that in typical scenarios these bounds are close and therefore allow one to accurately estimate the sum rate of the suggested schemes. The bounds also show that the developed schemes give significant improvement over other schemes in the literature (in particular the one given in [21]).

It is important to emphasize that we do not limit our study to only those systems with a large number of base-station antennas. We focus on such systems in the first part of the paper and develop simple precoder optimization that takes advantage of large number of base-station antennas. However, the design is applicable to systems with limited number of base-station antennas. In the second part of the paper, we study a modified version of the precoder presented in [22] that do not assume a large number of base-station antennas even for the design. In [22], a precoding matrix for downlink systems is obtained using an iterative algorithm which attempts to determine one of the local maxima of the sum rate maximization problem when CSI is available at the base-station and the users. Since, in our setting, the base-station obtains CSI through training and thus may not be perfect, we modify this algorithm to account for error in the estimation process.

### A. Prior Work

As is already well known, DPC [23] can be used as a precoding strategy when the interference signal is known non-causally and perfectly at the transmitter. Given that translating DPC to practice is by no means a trivial task, various alternative precoding methods with low complexity have been studied assuming perfect CSI. Prior work on precoding [24], [25], [26], [22], [27] demonstrates that sum rates close to sum capacity can be achieved with lower computational complexity compared to DPC. There are also opportunistic scheduling schemes [28] with lower complexity compared to DPC which can achieve sum rate that asymptotically scales identically as the sum capacity with the number of users. The existing literature on scheduling [29], [30] shows the significance of opportunistic scheduling towards maximizing the sum rate in the downlink.

As briefly mentioned before, in FDD systems, a limited-CSI setting has been studied in great detail primarily using a limited-feedback framework [15], [16], [17], [18], [19], [31], [32]. In this framework, perfect CSI is assumed at the users and limited-feedback to base-station is studied. In [17], the authors show that, at high SNR, the feedback rate required per user must grow linearly with the SNR (in dB) in order to obtain the full MIMO BC multiplexing gain. The main result in [18] is that CSI feedback can be significantly reduced by exploiting multi-user diversity. In [19], the authors design a joint CSI quantization, beamforming and scheduling algorithm to attain optimal throughput scaling. However, all these papers assume perfect channel knowledge at the users and do not study TDD systems. The effect of training in multi-user MIMO systems using TDD operation is studied in [21]. The authors limit the study to homogeneous users and zero-forcing precoding. Our paper is motivated from and builds on this work on TDD systems.

### B. Notation

We use bold face to denote vectors and matrices. All vectors are column vectors. We use $(\cdot)^T$ to denote the transpose, $(\cdot)^*$ to denote the conjugate and $(\cdot)^\dagger$ to denote the Hermitian of vectors and matrices. $\text{Tr}(\mathbf{A})$ denotes the trace of matrix $\mathbf{A}$ and $\mathbf{A}^{-1}$ denotes the inverse of matrix $\mathbf{A}$. $\text{diag}\{\mathbf{a}\}$ denotes a diagonal matrix with diagonal entries equal to the components of $\mathbf{a}$. $\succeq$ denotes element-wise greater than or equal to. $\mathbb{E}[\cdot]$ and $\text{var}\{\cdot\}$ stand for expectation and variance operations, respectively. $\mathbf{1}_{\{\cdot\}}$ denotes the indicator function.

### C. Organization

The rest of this paper is organized as follows. In Section II, we describe the system model and the assumptions. We consider two transmission methods. First, we consider a transmission method with channel training on reverse link only in Section III. Next, we consider a transmission method which sends forward pilots in addition to reverse pilots in Section IV. In Section V, we provide an upper bound on the sum rate for communication schemes using linear precoding at the base-station. We compare the performance of the various schemes considered through numerical results in Section VI and provide our concluding remarks in Section VII.

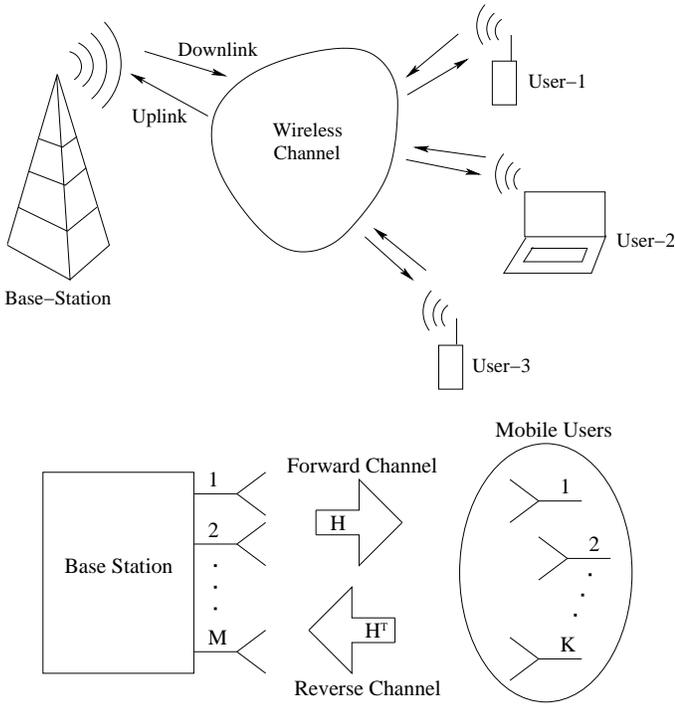

Fig. 1. Multi-user MIMO TDD system model

## II. SYSTEM MODEL

The system model consists of a base-station with $M$ antennas and $K$ single antenna users. The base-station communicates with the users on both forward and reverse links as shown in Figure 1. The forward channel is characterized by the $K \times M$ matrix $\mathbf{H}$ and the forward SNRs. The system model incorporates frequency selectivity of fading by using orthogonal frequency-division multiplexing (OFDM). The duration of the coherence interval (defined later) in symbols is chosen for one OFDM sub-band. For simplicity, we consider OFDM sub-bands as parallel channels and concentrate on one OFDM sub-band (where channel matrix is fixed and there is no multi-path). The details of OFDM (including cyclic prefix) are completely omitted, as this is by no means the focus of the paper. Further, we make the following assumptions.

1) Rayleigh block fading: The channel undergoes Rayleigh fading over blocks of $T$ symbols called the coherence interval during which the channel remains constant. In Rayleigh fading, the entries of the channel matrix $\mathbf{H}$ are independent and identically distributed (i.i.d.) zero-mean, circularly-symmetric complex Gaussian $\mathcal{CN}(0,1)$ random variables.
2) Reciprocity: The reverse channel between any user and the base-station (at any instant) is a scaled version of the forward channel.
3) Coherent uplink transmission: Time synchronization is present in the system.

Let the forward and reverse SNRs associated with $k$-th user be $\rho_k^f$ and $\rho_k^r$, respectively. These forward and reverse SNRs account for the average power at the base-station and the users, and the propagation factors (including path loss and shadowing). These propagation factors change at a much larger time-scale compared to fading. Hence, in the analysis, these parameters are treated as constants. For simplicity of notation, we ignore the time index. On the forward link, the signal received by the $k$-th user is

$$x_k^f = \sqrt{\rho_k^f}\, \mathbf{h}_k^T \mathbf{s}^f + z_k^f \tag{1}$$

where $\mathbf{h}_k^T$ is the $k$-th row of the channel matrix $\mathbf{H}$ and $\mathbf{s}^f$ is the $M \times 1$ signal vector. The additive noise $z_k^f$ is i.i.d. $\mathcal{CN}(0,1)$. The average power constraint at the base-station during transmission is $\mathbb{E}[\|\mathbf{s}^f\|^2] = 1$ so that the total transmit power is fixed irrespective of its number of antennas. The received power depends on the channel norm and hence on the number of antennas at the base-station. On the reverse link, the vector received at the base-station is

$$\mathbf{x}^r = \mathbf{H}^T \mathbf{E}^r \mathbf{s}^r + \mathbf{z}^r \tag{2}$$

where $\mathbf{s}^r$ is the signal-vector transmitted by the users and

$$\mathbf{E}^r = \mathrm{diag}\{[\sqrt{\rho_1^r}\ \sqrt{\rho_2^r}\ \cdots\ \sqrt{\rho_K^r}]^T\}.$$

The components of the additive noise vector $\mathbf{z}^r$ are i.i.d. $\mathcal{CN}(0,1)$. The power constraint at the $k$-th user during transmission is given by $\mathbb{E}[\|s_k^r\|^2] = 1$ where $s_k^r$ is the $k$-th component of $\mathbf{s}^r$.

*Remark 1:* We primarily focus on short coherence intervals. The need to study short coherence intervals arises from the high mobility of the users. In this setting, it is important that we account for channel training overhead and estimation error. Our goal is to account for these factors in the net throughput and develop schemes that achieve high net throughput. For obtaining schemes of practical importance, we look at schemes with low computational requirements. As mentioned earlier, we consider linear precoding techniques at the base-station.

*Remark 2:* The performance metric of interest is the achievable weighted-sum rate. The motivation behind looking at weighted-sum rate is that weights are used by higher layer protocols such as the Proportional Fair scheduling algorithm and the Max-Weight scheduling algorithm in order to achieve goals such as efficient fair sharing of throughput (Proportional Fair) and queue stabilization (Max Weight). For example, in the case of Max Weight they are fixed to be queue lengths [33]. The weights are passed to the physical layer, which has the task of maximizing the weighted-sum rate with given weights. It is this latter task and the performance achieved with which the paper is concerned. Thus, in a real system, these weights are adaptively controlled by the high-layer algorithm to perform a given network utility maximization [34].

By assumption, every user knows the system parameters such as the weights, the forward SNRs, the reverse SNRs and the achievable strategy. In typical systems, these parameters change on a much larger time-scale compared to the coherence interval and stays constant during many OFDM symbols. Typical shadow fading assumptions lead to the conclusion that significant SNR changes occur only over distance of 20 meters and above. Further, in communications standards like LTE, there are protocols that describe how SNRs are estimated and passed to base-stations. We do not address these in this paper, i.e., in our system model SNRs are assumed to be constant and known for the time-scale of interest. The symbol time of the LTE OFDM symbol is 71.3 $\mu$s. If a mobile moves with the speed of 50 miles/hour, then its SNR value will change after the transmission of approximately 12600 OFDM symbols whereas the channel coefficients will change within approximately 20 OFDM symbols. In a typical Proportional Fair algorithm, weights are kept fixed over a period of $1-10$ seconds. Hence, the number of OFDM symbols that will be transmitted in this time interval is again much larger than the coherence interval. These typical numbers clearly suggest that the overhead associated with learning system parameters is negligible compared to the channel training overhead, which is accounted for in this paper.

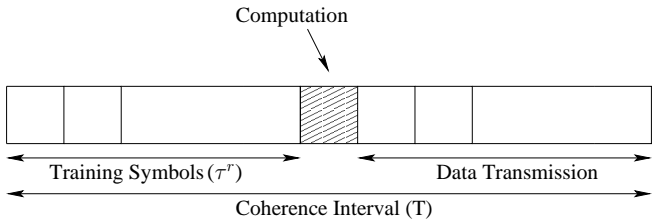

Fig. 2. Different phases in a coherence interval

## III. TRAINING ON REVERSE LINK ONLY

In this section, we consider a transmission scheme that consists of three phases as shown in Figure 2 - training, computation and data transmission. In the training phase, the users transmit training sequences to the base-station on the reverse link. The base-station performs the required computations for precoding in the computation phase. We assume that this causes a one-symbol delay in order to emphasize the delay in computation/control. In practice, this delay is a system dependent parameter. In the data transmission phase, the base-station transmits data symbols to the selected users.

*Remark 3:* In this transmission method, the users do not obtain any information regarding the instantaneous channel. The base-station obtains an estimate of the instantaneous channel. This is very different from the usual setting where the users also have estimates of channel gains. As a result, the analysis is very different as well.

Our goal is to obtain a simple precoding method that can achieve high weighted-sum rate. The capacity region of the system described in Section II is not known even in the single user setting. In addition, capacity achieving schemes can in general be very complex to implement in practice. Therefore, our approach is to obtain variants of well-studied simple algorithms in the perfect CSI setting that is applicable in the imperfect CSI setting, and analyze the system performance. In particular, we consider MMSE channel estimation, opportunistic selection of users based on channel gains, and generalized zero-forcing (described later) precoding. The parameters used in the algorithm are optimized for improved performance. The optimal precoding is identified in the course of an asymptotic analysis, taking the number of base-station antennas to infinity. Next, we provide the details of the algorithm and our analysis.

### A. Channel Estimation

Channel reciprocity is one of the key advantages of time-division duplex (TDD) systems over frequency-division duplex (FDD) systems. We exploit this property to perform channel estimation by transmitting training sequences on the reverse link. Every user transmits a sequence of training signals of $\tau^r$ symbols duration in every coherence interval. The $k$-th user transmits the training sequence vector $\sqrt{\tau^r}\,\boldsymbol{\psi}_k^\dagger$. We use orthonormal sequences which implies $\boldsymbol{\psi}_i^\dagger \boldsymbol{\psi}_j = \delta_{ij}$ where $\delta_{ij}$ is the Kronecker delta.

*Remark 4:* The use of orthogonal sequences restricts the maximum number of users to $\tau^r$, i.e., $K \leq \tau^r$.

The training signal matrix received at the base-station is

$$\mathbf{Y} = \sqrt{\tau^r}\,\mathbf{H}^T\mathbf{E}^r\boldsymbol{\Psi}^\dagger + \mathbf{V}^r$$

where $\boldsymbol{\Psi} = [\boldsymbol{\psi}_1\,\boldsymbol{\psi}_2\,\ldots\,\boldsymbol{\psi}_K]$ ($\boldsymbol{\Psi}^\dagger\boldsymbol{\Psi} = \mathbf{I}$) and the components of $\mathbf{V}^r$ are i.i.d. $\mathcal{CN}(0,1)$. The base-station obtains the linear minimum mean-square error estimate (LMMSE) of the channel

$$\hat{\mathbf{H}} = \mathrm{diag}\left\{\left[\frac{\sqrt{\rho_1^r\tau^r}}{1+\rho_1^r\tau^r}\,\cdots\,\frac{\sqrt{\rho_K^r\tau^r}}{1+\rho_K^r\tau^r}\right]^T\right\}\boldsymbol{\Psi}^T\mathbf{Y}^T. \quad (3)$$

The estimate $\hat{\mathbf{H}}$ is the conditional mean of $\mathbf{H}$ given $\mathbf{Y}$. Therefore, $\hat{\mathbf{H}}$ is the MMSE estimate as well. By the properties of conditional mean and joint Gaussian distribution, the estimate $\hat{\mathbf{H}}$ is independent of the estimation error $\tilde{\mathbf{H}} = \mathbf{H} - \hat{\mathbf{H}}$ [35]. The components of $\hat{\mathbf{H}}$ are independent and the elements of its $k$-th row are $\mathcal{CN}(0, \rho_k^r\tau^r/(1+\rho_k^r\tau^r))$. In addition, the components of $\tilde{\mathbf{H}}$ are independent and the elements of its $k$-th row are $\mathcal{CN}(0, 1/(1+\rho_k^r\tau^r))$.

### B. Generalized Zero-Forcing Precoding

Next, we describe a generalized zero-forcing (ZF) precoding. This precoding consists of two steps: (*i*) precoder parameter optimization, and (*ii*) user selection. The precoder parameters are non-negative constants $p_1,\ldots,p_K$, which are later optimized over long-term[1] system parameters such as the weights, the forward SNRs and the reverse SNRs. The user selection algorithm is denoted by $S(\hat{\mathbf{H}}) = \{S_1, S_2, \ldots, S_N\} \subseteq \{1, 2, \ldots, K\}$, i.e., based on the channel estimate $\hat{\mathbf{H}}$ the scheduling algorithm selects users $S_1, S_2, \ldots, S_N$. Thus, the user selection is dependent on short-term channel variations.

Before proceeding, we introduce the notation required to describe the precoding method. Let

$$\mathbf{D}_S = \mathrm{diag}\left\{\left[p_{S_1}^{-\frac{1}{2}}\,p_{S_2}^{-\frac{1}{2}}\,\ldots\,p_{S_N}^{-\frac{1}{2}}\right]^T\right\}.$$

Let $\hat{\mathbf{H}}_S$ be the $N \times M$ matrix formed from $\hat{\mathbf{H}}$ as follows: The $i$-th row ($1 \leq i \leq N$) corresponds to the $S_i$-th row of matrix $\hat{\mathbf{H}}$. Similarly, define $\mathbf{H}_S$ and $\tilde{\mathbf{H}}_S$. Let $\hat{\mathbf{H}}_{DS} = \mathbf{D}_S \hat{\mathbf{H}}_S$. Now, the generalized zero-forcing precoding matrix is defined as

$$\mathbf{A}_{DS} = \frac{\hat{\mathbf{H}}_{DS}^{\dagger} \left( \hat{\mathbf{H}}_{DS} \hat{\mathbf{H}}_{DS}^{\dagger} \right)^{-1}}{\sqrt{\text{Tr}\left[ \left( \hat{\mathbf{H}}_{DS} \hat{\mathbf{H}}_{DS}^{\dagger} \right)^{-1} \right]}}. \quad (4)$$

---

[1] Strictly speaking, these long-term parameters are constants in our system model.

This precoding matrix is normalized so that

$$\text{Tr}\left( \mathbf{A}_{DS}^{\dagger} \mathbf{A}_{DS} \right) = 1.$$

The matrix $\mathbf{D}_S$ is introduced to optimally allocate "resources" to users. This is required as our system consists of heterogeneous users.

Let $\mathbf{q}$ denote the vector of (coded) information symbols that have to transmitted to the $N$ selected users. Then, the transmission signal-vector is given by

$$\mathbf{s}_f = \mathbf{A}_{DS} \mathbf{q}. \quad (5)$$

Clearly, the base-station transmit power constraint can be satisfied irrespective of the values of $p_1, \ldots, p_K$ by imposing the conditions $\mathbb{E}[\|q_n\|^2] = 1, \forall n \in \{1, \ldots, N\}$.

This generalized zero-forcing precoding method requires a choice of the $p_i$ values and a user selection algorithm. Next, we characterize the achievable throughput with this precoding method, and then explain the precoder optimization and the user selection algorithm.

### C. Achievable Throughput

In this section, we obtain an achievable throughput for the system under consideration (by building on techniques in [36]). Given a user selection algorithm, we denote the probability of selecting the $k$-th user as $\gamma_k$. The throughput derived depends on the user selection strategy through the random variable $\chi$ (defined later) and the probabilities of selecting the users. Recall that $M$ is the number of antennas at the base-station, $K$ is the number of users, $\rho_k^f$ is the forward SNR associated with the $k$-th user and $\rho_k^r$ is the reverse SNR associated with the $k$-th user. Let the weight associated with the $k$-th user be $w_k$. The base-station performs MMSE channel estimation as described in Section III-A. For channel estimation, the training period used is $\tau^r \geq K$ symbols.

From (1), the signal-vector received at the selected users (according to our system model the user knows whether it is selected or not) is

$$\mathbf{x}^f = \mathbf{E}_S^f \mathbf{H}_S \mathbf{A}_{DS} \mathbf{q} + \mathbf{z}^f \quad (6)$$

where

$$\mathbf{E}_S^f = \text{diag}\left\{ \left[ \sqrt{\rho_{S_1}^f} \ \sqrt{\rho_{S_2}^f} \ \cdots \ \sqrt{\rho_{S_N}^f} \right]^T \right\}.$$

The effective forward channel in (6) is

$$\begin{aligned}
\mathbf{G} &= \mathbf{E}_S^f \mathbf{H}_S \mathbf{A}_{DS} \\
&= \mathbf{E}_S^f \left( \mathbf{D}_S^{-1} \hat{\mathbf{H}}_{DS} + \tilde{\mathbf{H}}_S \right) \mathbf{A}_{DS} \\
&= \mathbf{E}_S^f \mathbf{D}_S^{-1} \chi + \mathbf{E}_S^f \tilde{\mathbf{H}}_S \mathbf{A}_{DS}, \quad (7)
\end{aligned}$$

where $\chi$ is the scalar random variable given by

$$\chi = \left( \text{Tr}\left[ \left( \hat{\mathbf{H}}_{DS} \hat{\mathbf{H}}_{DS}^{\dagger} \right)^{-1} \right] \right)^{-\frac{1}{2}}. \quad (8)$$

Suppose that the $k$-th user is among the selected users. The signal received by the $k$-th user is

$$x_k^f = \mathbf{g}^T \mathbf{q} + z_k^f \quad (9)$$

where $\mathbf{g}^T$ is the row corresponding to $k$-th user in matrix $\mathbf{G}$. From (7), we obtain

$$\mathbf{g}^T = \sqrt{\rho_k^f p_k}\, \chi \mathbf{e}_k^T + \sqrt{\rho_k^f}\, \tilde{\mathbf{h}}_k^T \mathbf{A}_{DS} \quad (10)$$

where $\tilde{\mathbf{h}}_k^T$ is the $k$-th row of $\tilde{\mathbf{H}}$ and $\mathbf{e}_k$ is the $N \times 1$ column-vector with $k$-th element equal to one and all other elements equal to zero. Substituting (10) in (9) and adding and subtracting mean from $\chi$, we obtain

$$\begin{aligned}
x_k^f &= \sqrt{\rho_k^f p_k}\, \mathbb{E}[\chi]\, q_k + \sqrt{\rho_k^f p_k}\, (\chi - \mathbb{E}[\chi])\, q_k \\
&\quad + \sqrt{\rho_k^f}\, \tilde{\mathbf{h}}_k^T \mathbf{A}_{DS} \mathbf{q} + z_k^f \quad (11) \\
&= \sqrt{\rho_k^f p_k}\, \mathbb{E}[\chi]\, q_k + \hat{z}_k^f
\end{aligned}$$

where the effective noise

$$\hat{z}_k^f = \sqrt{\rho_k^f p_k}\, (\chi - \mathbb{E}[\chi])\, q_k + \sqrt{\rho_k^f}\, \tilde{\mathbf{h}}_k^T \mathbf{A}_{DS} \mathbf{q} + z_k^f.$$

According to our system model, each user knows the systems parameters. However, the users do not know the instantaneous channels, which is the main overhead that is often neglected. Hence, the user performs the following:

1) It computes the expected value (over instantaneous channel distribution) of its "effective" channel given by $(\rho_k^f p_k)^{1/2} \mathbb{E}[\chi]$. In other words, this is the expected gain multiplying its information symbol.
2) It computes the variation of the effective channel around its expected value given by $\rho_k^f p_k \textbf{var}\{\chi\}$. This contributes to the "effective" noise variance.
3) It computes remaining terms that contribute to effective noise variance, which includes the interference due to other information signals given by $\rho_k^f/(1 + \rho_k^r \tau^r)$ and the additive noise variance (which is unity).
4) It computes the effective SNR from the above computations, and uses it in the decoding.

In the following theorem, we formalize the above by showing that the effective noise is uncorrelated with signal and use this fact to obtain achievable weighted-sum rate.

*Theorem 1:* Consider the precoding method described above. Then, the following weighted-sum rate is achievable during downlink transmission:

$$R_\Sigma = \sum_{k=1}^{K} \gamma_k w_k \log_2 \left( 1 + \frac{\rho_k^f p_k \mathbb{E}^2[\chi]}{1 + \rho_k^f \left( \frac{1}{1 + \rho_k^r \tau^r} + p_k \textbf{var}\{\chi\} \right)} \right), \quad (12)$$

where $\chi$ is the scalar random variable in (8).

*Proof:* The expected value of any term on the right-hand side of (11) is zero. The noise term $z_k^f$ is independent of all other terms and

$$\mathbb{E}\left[ z_k^f \Big| \mathbf{q} \right] = 0, \quad \mathbb{E}\left[ z_k^f \Big| \mathbf{q}, \hat{\mathbf{H}} \right] = 0, \quad \mathbb{E}\left[ \tilde{\mathbf{h}}_k^T \Big| \mathbf{q}, \hat{\mathbf{H}} \right] = 0.$$

Using the law of iterated expectations, we have

$$\mathbb{E}\left[q_k q_k^\dagger \left(\chi - \mathbb{E}\left[\chi\right]\right)\right] = \mathbb{E}\left[q_k q_k^\dagger\right]\left(\mathbb{E}\left[\chi\right] - \mathbb{E}\left[\chi\right]\right) = 0,$$

$$\mathbb{E}\left[q_k \mathbf{q}^\dagger \mathbf{A}_{DS}^\dagger \tilde{\mathbf{h}}_k^*\right] = \mathbb{E}\left[q_k \mathbf{q}^\dagger \mathbf{A}_{DS}^\dagger \mathbb{E}\left[\tilde{\mathbf{h}}_k^* \Big| \mathbf{q}, \hat{\mathbf{H}}\right]\right] = 0,$$

$$\mathbb{E}\left[\left(\chi - \mathbb{E}\left[\chi\right]\right) q_k \mathbf{q}^\dagger \mathbf{A}_{DS}^\dagger \tilde{\mathbf{h}}_k^*\right] =$$
$$\mathbb{E}\left[\left(\chi - \mathbb{E}\left[\chi\right]\right) q_k \mathbf{q}^\dagger \mathbf{A}_{DS}^\dagger \mathbb{E}\left[\tilde{\mathbf{h}}_k^* \Big| \mathbf{q}, \hat{\mathbf{H}}\right]\right] = 0.$$

Hence, any two terms on the right-hand side of (11) are uncorrelated. The effective noise $\hat{z}_k^f$ is thus uncorrelated with the signal $q_k$. The effective noise has zero mean and variance

$$\begin{aligned}\mathbf{var}\left\{\hat{z}_k^f\right\} &= 1 + \rho_k^f \mathbb{E}\left[\tilde{\mathbf{h}}_k^T \mathbf{A}_{DS} \mathbb{E}\left[\mathbf{q}\mathbf{q}^\dagger \Big| \hat{\mathbf{H}}, \tilde{\mathbf{H}}\right] \mathbf{A}_{DS}^\dagger \tilde{\mathbf{h}}_k^*\right] \\ &\quad + \rho_k^f p_k \mathbf{var}\left\{\chi\right\} \\ &= 1 + \rho_k^f \left(\frac{1}{1+\rho_k^r \tau^r} + p_k \mathbf{var}\left\{\chi\right\}\right).\end{aligned}$$

*Remark 5:* The effective noise $\hat{z}_k^f$ is uncorrelated with the signal $q_k$, and in general not independent. Note that we do not need independence in the proof.

In order to obtain a set of achievable rates, we consider $(T - \tau^r - 1)$ parallel channels where noise is independent over time as fading is independent over blocks. Using the fact that worst-case uncorrelated noise distribution is independent Gaussian noise with same variance, we obtain the achievable weighted-sum rate given in (12). This completes the proof. ∎

The proof assumes that the users know if they are selected or not. In Section III-E, we discuss how this assumption can be relaxed with a small reduction in net achievable rate.

*Remark 6:* The values $\mathbb{E}[\chi]$ and $\mathbf{var}\{\chi\}$ do not depend on short-term channel variations. $\mathbb{E}[\chi]$ and $\mathbf{var}\{\chi\}$ depend only on slowly changing parameters, namely on the weights, the reverse SNRs and the user selection strategy. These slowly changing parameters stay constant over a large period comprising many coherence intervals. We assume that these parameters are known at the base-station and corresponding users. The values $\mathbb{E}[\chi]$ and $\mathbf{var}\{\chi\}$ can be accurately estimated via a Monte-Carlo simulation in the beginning of each period. These estimates can be produced either by users themselves or by the base-stations. In the latter case, the base-station will have to pass the values $\mathbb{E}[\chi]$ and $\mathbf{var}\{\chi\}$ to the corresponding users, which would assume only a small overhead. Alternatively, one can generate a look up table for $\mathbb{E}[\chi]$ and $\mathbf{var}\{\chi\}$ for a grid of parameter values. For intermediate cases, the corresponding values can be found by interpolation.

### D. Optimization of Precoding Matrix

We introduced the parameters $p_1, \ldots, p_K$ in the generalized zero-forcing precoding to handle the heterogeneity of users, i.e., differences in the weights, the forward SNRs and the reverse SNRs associated with users. In this section, our goal is to obtain these parameters as a function of the weights, the forward SNRs and the reverse SNRs. We make the following simplifications to achieve our goal.

1) The performance metric of interest is the achievable weighted-sum rate $R_\Sigma$ in (12). However, $R_\Sigma$ is a function of the user selection algorithm. To overcome this, we simply consider the case of selecting all users to obtain $p_1, \ldots, p_K$. Hence, this can be performed before the user selection.
2) We would like to choose non-negative values for $p_1, \ldots, p_K$ such that $R_\Sigma$ in (12) is maximized. However, this is a hard problem to analyze as closed-form expression for the expectation and the variance terms in (12) is unknown. We consider the asymptotic regime $M/K \gg 1$ as this is appropriate in this section.

*Remark 7:* Apart from making the problem mathematically tractable, the asymptotic regime $M/K \gg 1$ is of interest due to the following reasons: (*i*) the system constraints $K \leq \tau^r$, $\tau^r \leq T$ place an upper bound on $K$, independent of the number of antennas, and (*ii*) the base-station can be equipped with many antennas each powered by its own low-power tower-top amplifier [21].

From the weak law of large numbers, it is known that

$$\lim_{M/K \to \infty} \frac{1}{M} \mathbf{Z}\mathbf{Z}^\dagger = \mathbf{I}_K$$

where $\mathbf{Z}$ is the $K \times M$ random matrix whose elements are i.i.d. $\mathcal{CN}(0, 1)$. Therefore, $\mathbf{Z}\mathbf{Z}^\dagger$ can be approximated by $M\mathbf{I}_K$. Hence, the random variable $\chi$ in (8) can be approximated as

$$\chi \approx \sqrt{\frac{M}{\sum\limits_{j=1}^{K} a_j p_j}} \tag{13}$$

where

$$a_j = \left(\frac{\rho_j^r \tau^r}{1 + \rho_j^r \tau^r}\right)^{-1}.$$

Substituting (13) in (12), we get

$$R_\Sigma \approx J(\mathbf{p}) = \sum_{i=1}^{K} w_i \log_2\left(1 + \frac{b_i p_i}{\sum\limits_{j=1}^{K} a_j p_j}\right)$$

where

$$b_i = \frac{M\rho_i^f}{1 + \rho_i^f (1 + \rho_i^r \tau^r)^{-1}}.$$

Under this approximation, we can find the optimal values for $p_1, \ldots, p_K$ that maximize $J(\mathbf{p})$ as described below.

*Theorem 2:* An optimal solution $\mathbf{p}^*$ of the objective function $\max_{\mathbf{p}} J(\mathbf{p})$ is of the form $c\overline{\mathbf{p}}^*$ where $c$ is any positive real number and $\overline{\mathbf{p}}^* = [\overline{p}_1^* \ \overline{p}_2^* \ \ldots \ \overline{p}_K^*]^T$ is given by

$$\overline{p}_i^* = \max\left\{0, \left(\frac{w_i}{\nu^* a_i} - \frac{1}{b_i}\right)\right\}. \tag{14}$$

The positive real number $\nu^*$ is unique and given by

$$\sum_{i=1}^{K} a_i \overline{p}_i^* = 1.$$

*Proof:* The proof idea is to introduce an additional constraint to obtain a convex optimization problem. We show that the introduction of the additional constraint does not affect the optimal value of the optimization problem.

Note that $w_i > 0$, $b_i > 0$ and $a_j > 0$. Let $\mathbf{a} = [a_1 \, a_2 \, \ldots \, a_K]^T$. We consider the optimization problem

$$\text{maximize} \quad J(\mathbf{p}) \tag{15}$$
$$\text{subject to} \quad \mathbf{p} \succeq 0.$$

Since $J(\mathbf{p}) = J(c\mathbf{p})$ for any $c > 0$ and $\mathbf{p}^* \neq 0$, $\mathbf{p}^*$ such that $\mathbf{a}^T\mathbf{p}^* = c$ is an optimal solution to (15) if and only if $\overline{\mathbf{p}}^* = (1/c)\mathbf{p}^*$ is an optimal solution to the convex optimization problem

$$\text{minimize} \quad -\sum_{i=1}^{K} w_i \log\left(1 + b_i \overline{p}_i\right) \tag{16}$$
$$\text{subject to} \quad \overline{\mathbf{p}} \succeq 0, \mathbf{a}^T \overline{\mathbf{p}} = 1.$$

In order to solve (16), we introduce Lagrange multipliers $\boldsymbol{\lambda} \in \mathbb{R}^K$ for the inequality constraints $\overline{\mathbf{p}} \succeq 0$ and $\nu \in \mathbb{R}$ for the equality constraint $\mathbf{a}^T \overline{\mathbf{p}} = 1$. The necessary and sufficient conditions for optimality are given by Karush-Kuhn-Tucker (KKT) conditions [37]. These conditions are

$$\overline{\mathbf{p}}^* \succeq 0, \quad \mathbf{a}^T \overline{\mathbf{p}}^* = 1, \quad \boldsymbol{\lambda}^* \succeq 0,$$
$$\lambda_i^* \overline{p}_i^* = 0, \quad -\frac{w_i b_i}{1 + b_i \overline{p}_i^*} - \lambda_i^* + \nu^* a_i = 0, \quad i = 1, \ldots, K.$$

This set of equations can be simplified to

$$\overline{p}_i^* = \max\left\{0, \left(\frac{w_i}{\nu^* a_i} - \frac{1}{b_i}\right)\right\},$$
$$\sum_{i=1}^{K} a_i \max\left\{0, \left(\frac{w_i}{\nu^* a_i} - \frac{1}{b_i}\right)\right\} = 1. \tag{17}$$

Since the left-hand of (17) is an increasing function in $1/\nu^*$, this equation has a unique solution, which can be easily computed numerically using binary search. This completes the proof. ∎

The optimized $\overline{\mathbf{p}}^*$ given by (14) is substituted in (4) to obtain the optimized precoding matrix. We remark that this precoder design is only asymptotically optimal when $M/K \to \infty$. However, we use this optimized precoding matrix even when number of users $K$ is comparable to number of base-station antennas $M$. We denote the scheme where we use optimized $p_i$ values for precoding by Scheme-1 and the scheme where we use $p_i = 1$ for precoding by Scheme-0. In both the schemes, we select all the users.

### E. User Selection Strategy

We consider a simple user selection strategy based on opportunistic selection of users based on scaled estimated channel gains of users (details given later). We ignore the spatial separability/orthogonality of channels due to the following reason. As mentioned earlier, the transmission method in this section is of interest in the large number of base-station antennas setting. In this setting, the spatial separability/orthogonality of channel play a less important role. Also, the channel estimate at the base-station is expected to be poor. The prediction of channel orthogonality based on this poor estimate is generally inaccurate. In addition, brute-force search over subsets of users is computationally complex. In the second part of this paper, for the general setting, we consider schemes that use spatial separability/orthogonality of channels.

In the user selection strategies presented below, we need not assume any designated channel for informing users whether they were selected or not. We will show that this does not result in a significant loss of data rate.

Let us consider a selection scheme with a designated channel for alerting the selected users. Let $I_j$ be the average amount of information (in bits) that can be transmitted to the $j$-th user during one coherence interval in this scheme. The averaging is conducted over multiple coherence intervals in which the user can be selected or not. Denote by $I'_j$ the corresponding quantity for the same user selection scheme, but without the designated channel.

*Theorem 3:*
$$I'_j \geq I_j - 1.$$

*Proof:* Let $D_j$ be the random variable indicating whether or not the $j$-th user was selected in a given coherence interval. The value $D_j = 1$ indicates that the user was selected and $D_j = 0$ indicates that it was not. Let $A_{D_1,\ldots,D_K}$ be a precoding matrix. This matrix depends on the values $D_1, \ldots, D_K$. In particular the $j$-th column of $A_{D_1,\ldots,D_K}$ is the all-zero vector if $D_j = 0$. Denote by $q_j^t$ the symbol that is transmitted to the $j$-th user at time instance $t$. Then, the signal received by the $j$-th user is

$$x_j^t = \sqrt{\rho_j^f} \mathbf{h}_j^T A_{D_1,\ldots,D_K} \begin{pmatrix} q_1^t \\ \vdots \\ q_K^t \end{pmatrix} + z_k^t, \tag{18}$$

where $z_k^t$ is Gaussian noise. Denote $\mathbf{x}_j = (x_j^{\tau^r+2}, \ldots, x_j^T)$ and $\mathbf{q}_j = (q_j^{\tau^r+2}, \ldots, q_j^T)$. If the designated channel is available, we have

$$I_j = I(\mathbf{x}_j; \mathbf{q}_j D_j).$$

It is important to note that this mutual information is over the communication channel defined by (18), which includes not only the MIMO transmission, but also the random variables $D_j$. Note also that $D_j$ and $\mathbf{q}_j$ are independent. Similarly, if the designated channel is absent, we have

$$I'_j = I(\mathbf{x}_j; \mathbf{q}_j).$$

Using the chain rule for mutual information, we obtain

$$I_j = I(\mathbf{x}_j; \mathbf{q}_j) + I(\mathbf{x}_j; D_j | \mathbf{q}_j) = I'_j + I(\mathbf{x}_j; D_j | \mathbf{q}_j).$$

Since $D_j$ is a binary random variable, we have

$$I(\mathbf{x}_j; D_j | \mathbf{q}_j) \leq 1,$$

and the assertion follows. ∎

It is worth noting that applying the chain rule, we obtain

$$\begin{aligned} I_j &= I(\mathbf{q}_j; D_j) + I(\mathbf{x}_j; \mathbf{q}_j | D_j) = I(\mathbf{x}_j; \mathbf{q}_j | D_j) \\ &= I_{\text{MIMO}}(\mathbf{x}_j; \mathbf{q}_j) \Pr(D_j = 1). \end{aligned} \tag{19}$$

These equalities follows from the fact that $\mathbf{q}_j$ and $D_j$ are independent random variables and from the fact $I(\mathbf{x}_j; \mathbf{q}_j | D_j =$

0) = 0. In (19), the mutual information $I_{\text{MIMO}}(\mathbf{x}_j; \mathbf{q}_j)$ is over the channel defined in (9), that is over the MIMO (the dimensions are time, not antennas) channel for *selected users*, which is different from the channel (18).

Another important concern is the practical realization of the user selection scheme. One possible way is to design a criterion that would allow each user to decide whether it was selected or not for each coherence interval. For instance, one may try to use the power of the received signal $\mathbf{x}_j$ as such criterion. We believe that this is a poor approach, which incorporates a hard decision, which results in rate loss.

A significantly better way is to assume the channel model (18) in which we always "transmit" signals $q_j$ independent on whether the $j$-th user was selected or not. This is equivalent to data transmission via a fading channel of the form $x_j = const \cdot q_j D_j + noise$. For recovering the transmitted symbols, we propose to use an error correcting code, which approaches the capacity of this fading channel. In contrast to the previous hard decision approach, the probabilities $\Pr(D_j = 0|x_j), \Pr(D_j = 1, q_j|x_j)$ for all possible values of $q_j$, and passed to the decoder. It is not difcult to construct an LDPC code approaching capacity of this fading channel, using, for instance, the EXIT function technique described in [38], [39]. A decoder of such an LDPC code will update the probabilities $\Pr(D_j = 0|x_j), \Pr(D_j = 1, q_j|x_j)$ using intermediate decoding results from each iteration (see [39] for details of this technique).

*1) Homogeneous Users:* First, we consider the special case where the users are statistically identical. In this homogeneous setting, the forward SNRs from the base-station to all the users are equal (given by $\rho^f$) and reverse SNRs from all the users to the base-station are equal (given by $\rho^r$). Furthermore, the weights assigned to all the users are unity, i.e., $w_k = 1$. The need for explicit user selection arises due to the ZF based precoding used. With perfect channel knowledge at the base-station ($\hat{\mathbf{H}} = \mathbf{H}$) and no user selection ($N = K$), the ZF precoding diagonalizes the effective forward channel and all users see same effective channel gains.

We use the following simple heuristic rule at the base-station. In every coherence interval, the base-station selects those $N$ users with largest estimated channel gains. This rule is motivated by the expectation term $\mathbb{E}[\chi]$ appearing in the achievable weighted-sum rate in (12). Let $\hat{\mathbf{h}}_{(1)}^T, \hat{\mathbf{h}}_{(2)}^T, \ldots, \hat{\mathbf{h}}_{(K)}^T$ be the norm-ordered rows of the estimated channel matrix $\hat{\mathbf{H}}$. Then, the matrix $\hat{\mathbf{H}}_S$ is given by $\hat{\mathbf{H}}_S = [\hat{\mathbf{h}}_{(1)} \hat{\mathbf{h}}_{(2)} \ldots \hat{\mathbf{h}}_{(N)}]^T$ and the achievable sum rate in (12) becomes

$$R_\Sigma = N \log_2 \left( 1 + \frac{\rho^f \left( \frac{\rho^r \tau^r}{1+\rho^r \tau^r} \right) \mathbb{E}^2[\eta]}{1 + \rho^f \left( \frac{1}{1+\rho^r \tau^r} + \frac{\rho^r \tau^r}{1+\rho^r \tau^r} \mathbf{var}\{\eta\} \right)} \right). \quad (20)$$

Here, the random variable

$$\eta = \left( \text{Tr} \left[ (\mathbf{U}\mathbf{U}^\dagger)^{-1} \right] \right)^{-\frac{1}{2}}$$

where $\mathbf{U}$ is the $N \times M$ matrix formed by the $N$ rows with largest norms of a $K \times M$ random matrix $\mathbf{Z}$ whose elements are i.i.d. $\mathcal{CN}(0, 1)$.

Net achievable sum rate accounts for the reduction in achievable sum rate due to training. In every coherence interval of $T$ symbols, first $\tau^r$ symbols are used for training on reverse link, a symbol is used for computation and the remaining $(T - \tau^r - 1)$ symbols are used for transmitting information symbols as shown in Figure 2. The training length $\tau^r$ can be chosen such that net throughput of the system is maximized. For $N < K$, the sum rate overhead associated with user selection would be $K/T$. Thus, the net achievable sum rate is defined as

$$R_{\text{net}} = \max_{\tau^r, N} \left( \frac{T - \tau^r - 1}{T} R_\Sigma - \mathbf{1}_{\{N<K\}} \frac{\sum_{i=1}^{K} w_i}{T} \right) \quad (21)$$

subject to $N \leq K$, $\tau^r \leq T - 1$ and $\tau^r \geq K$. In (21), we optimize over both $N$ and $\tau^r$.

*2) Heterogeneous Users:* In this section, we consider the following heuristic user selection strategy for heterogeneous users.

Let $\mathbf{z}_1^T, \mathbf{z}_2^T, \ldots, \mathbf{z}_K^T$ be the rows of the matrix

$$\mathbf{Z} = \text{diag} \left\{ \left[ \sqrt{\frac{1+\rho_1^r \tau^r}{\rho_1^r \tau^r}} \cdots \sqrt{\frac{1+\rho_K^r \tau^r}{\rho_K^r \tau^r}} \right]^T \right\} \hat{\mathbf{H}}$$

where $\hat{\mathbf{H}}$ is the estimated channel given by (3). Note that $\mathbf{Z}$ is normalized such that the entries are independent and identically distributed. In every coherence interval, the users are ordered such that

$$\overline{p}_{(1)}^* \|\mathbf{z}_{(1)}^T\|^2 \geq \overline{p}_{(2)}^* \|\mathbf{z}_{(2)}^T\|^2 \geq \ldots \geq \overline{p}_{(K)}^* \|\mathbf{z}_{(K)}^T\|^2$$

and the first $N$ users under this ordering are selected. The value $N_{opt}$ is used for $N$ that maximize the net achievable weighted-sum rate defined below. The intuition behind this strategy is that $\overline{p}_{(k)}^*$ is nearly proportional to the average power assigned to the $k$-th user and $\|\mathbf{z}_{(k)}^T\|^2$ captures the instantaneous variation in power. Similar to the homogeneous case, the net achievable weighted-sum rate is given by (21). We denote the scheme where we use this user selection strategy along with optimized $p_i$ values for precoding by Scheme-2. We provide numerical results showing the improvement obtained by using this strategy in Section VI.

### F. Optimal Training Length

We consider the problem of finding the optimal training length in the homogeneous setting when the user selection strategy given in Section III-E is used. The objective is to maximize the net achievable sum rate given by (21). For given values of $M, K, T, \rho^f$ and $\rho^r$, it seems intractable to obtain a closed-form expression for the optimal training length. Therefore, we look at the limiting cases $\rho^r \to 0$ and $\rho^r \to \infty$ to understand the behavior of the optimal training length with reverse SNR.

In the limit $\rho^r \to 0$, we can approximate the net rate as

$$R_{\text{net}} \approx \frac{T - \tau^r - 1}{T} N \log_2 \left( 1 + \frac{\rho^f \rho^r \tau^r}{1 + \rho^f} \mathbb{E}^2[\eta] \right).$$

We use the fact that $\log(1+x) \approx x$ as $x \to 0$ to obtain the approximation

$$R_{\text{net}} \approx d_1 \frac{T - \tau^r - 1}{T} \tau^r \tag{22}$$

where $d_1$ is a positive constant. It is clear that (22) is maximized when $\tau^r = (T-1)/2$ if we assume $T > 2K$ and $T$ is odd. In the limit $\rho^r \to \infty$, we can approximate the net rate as

$$R_{\text{net}} \approx d_2 \frac{T - \tau^r - 1}{T}$$

where $d_2$ is a positive constant. This expression is maximized by the minimum possible training length which is $\tau^r = K$.

The approximations suggests that nearly half the coherence time should be spent for training when the reverse SNR is very low and the minimum possible number of symbols (which is $K$) should be spent for training when reverse SNR is very high. This conclusion is similar to the result in [36] for MIMO.

In summary, we developed a new precoding method referred to as generalized zero-forcing precoding. It consists of a user selection component and an optimization component. The user selection component is performed using opportunistic selection heuristics. The optimization component is performed using a convex optimization problem resulting from a relevant asymptotics of large number of base-station antennas. The resulting precoding is simple and therefore has significant practical value. We demonstrate the improvement obtained in net throughput through numerical examples in Section VI.

The net throughput improvement results from all optimization parameters. The role of the training length parameter is clear, as there is tension between large training overhead and better channel estimation. The more subtle parameters are the number of selected users $N$ and the precoder parameters $p_1, \ldots, p_K$. The role of the precoder parameters is to take advantage of long-term system parameters and statistics such as the weights, the forward SNRs and the reverse SNRs whereas the role of the parameter $N$ is take advantage of the short-term channel variations. In our approach, since the precoder optimization is dependent on long-term variations, it is not dependent on $N$. The choice of $N$ would depend on precoder parameters and is therefore more involved. However, since it is a single parameter, this optimization can be handled.

## IV. TRAINING ON REVERSE AND FORWARD LINKS

In this section, we consider a transmission method which sends forward pilots in addition to reverse pilots in Section IV[2]. In this section, we do not limit our approach to large number of base-station antennas.

In the transmission method considered in the previous section, the users do not obtain any knowledge about the instantaneous channel. Every user can be provided with partial knowledge about its effective channel gain in one of the following two ways. (*i*) The base-station can send quantized information of the effective channel gains to the users. (*ii*) The base-station can send forward pilots to the users so that the users can estimate the effective gains. It is hard to account for the overhead when base-station send quantized information about the effective channel gains. In addition, pilot based channel training is conventional in wireless systems. Therefore, we focus on sending pilots in the forward link. This leads to a transmission method consisting of four phases - reverse pilots, computation phase, forward pilots and data transmission - as shown in Figure 3. In this scheme, the users can obtain effective channel gain estimates at the expense of increased training overhead.

---

[2]There has been some parallel work in [40]. The authors consider two-way training [41] and study two variants of linear MMSE precoders as alternatives to linear zero-forcing precoder used in [21].

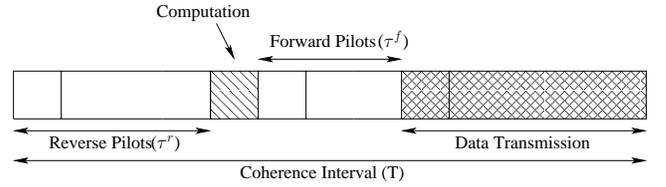

Fig. 3.  Reverse and Forward Pilots

### A. Channel Estimation and Precoding

As explained in Section III-A, the users transmit orthogonal training sequences on the reverse link. From these training sequences, the base-station obtains the MMSE estimate of the channel. The base-station uses this channel estimate $\hat{\mathbf{H}}$ to form a precoding matrix to perform linear precoding. Let $\mathbf{A}$ denote any precoding matrix which is a function of the channel estimate, i.e., $\mathbf{A} = f(\hat{\mathbf{H}})$. The precoding function $f(\cdot)$ usually depends on the system parameters such as forward SNRs, reverse SNRs and weights assigned to the users. We require that the precoding matrix is normalized so that $\text{Tr}\left(\mathbf{A}^\dagger \mathbf{A}\right) = 1$. The transmission signal-vector is given by $\mathbf{s}_f = \mathbf{A}\mathbf{q}$, where $\mathbf{q} = [q_1 \, q_2 \, \ldots \, q_K]^T$ is the vector of information symbols for the users. The net achievable rate derived later in this section is valid for any precoding function. Next, we describe a particular precoding method.

In [22], the following approach was suggested for finding a good precoding matrix $\mathbf{A}$. Let $\mathbf{h}_i$ be the $i$-th row of the channel matrix $\mathbf{H}$ and let $\mathbf{a}_j$ be the $j$-th column of precoding matrix $\mathbf{A}$. The sum rate of the broadcast channel can be written in the form

$$R(\mathbf{H}, \mathbf{A}) = \sum_{j=1}^{M} \log_2 \left(1 + \frac{|\mathbf{h}_j \mathbf{a}_j|^2}{\sigma^2 \text{Tr}\left(\mathbf{A}\mathbf{A}^\dagger\right) + \sum_{l \neq j} |\mathbf{h}_j \mathbf{a}_l|^2}\right).$$

Let

$$b_j = |\mathbf{h}_j \mathbf{a}_j|^2 \text{ and } c_j = \sigma^2 \text{Tr}\left(\mathbf{A}\mathbf{A}^\dagger\right) + \sum_{l \neq j} |\mathbf{h}_j \mathbf{a}_l|^2.$$

Further, let $\mathbf{\Delta}$ and $\mathbf{D}$ be diagonal matrices defined as

$$\mathbf{\Delta} = \text{diag}\left\{\left[\frac{(\mathbf{HA})_{11}}{c_1} \, \frac{(\mathbf{HA})_{22}}{c_2} \, \ldots \, \frac{(\mathbf{HA})_{MM}}{c_M}\right]^T\right\} \tag{23}$$

and

$$\mathbf{D} = \text{diag}\left\{\left[\frac{b_1}{c_1(b_1 + c_1)} \, \ldots \, \frac{b_M}{c_M(b_M + c_M)}\right]^T\right\}. \tag{24}$$

10In [22], it is shown that the equations $\frac{\partial R(\mathbf{H},\mathbf{A})}{\partial \mathbf{A}_{ij}} = 0$ imply

$$\mathbf{A} = ((\sigma^2 \text{Tr}(\mathbf{D}))I_M + \mathbf{H}^\dagger \mathbf{D}\mathbf{H})^{-1}\mathbf{H}^\dagger \mathbf{\Delta}. \quad (25)$$

This equation allows one to use the following iterative algorithm for determining an efficient $\mathbf{A}$:

1) Assign some initial values to matrices $\mathbf{\Delta}$ and $\mathbf{D}$, for instance $\mathbf{\Delta} = I_M, \mathbf{D} = I_M$
2) Repeat steps 3 and 4 several times
3) Compute $\mathbf{A}$ according to (25);
4) Compute $\mathbf{\Delta}$ and $\mathbf{D}$ according to (23) and (24).

This approach can be extended for the scenario when only an estimate $\hat{\mathbf{H}}$ of the channel matrix $\mathbf{H}$ and the statistics of the estimation error $\tilde{\mathbf{H}}$ is available. In this case, we would like to maximize the value of the average sum rate defined by

$$R(\hat{\mathbf{H}}, \mathbf{A}) = \mathbb{E}_{\tilde{\mathbf{H}}}[R(\hat{\mathbf{H}} + \tilde{\mathbf{H}}, \mathbf{A})].$$

Since the statistics of $\tilde{\mathbf{H}}$ is assumed to be known, we can generate $L$ samples $\tilde{\mathbf{H}}^{(i)}, i = 1, \ldots, L$, according to the statistics. Define $\mathbf{H}^{(i)} = \hat{\mathbf{H}} + \tilde{\mathbf{H}}^{(i)}$. Then, the average rate can be approximated as $R(\hat{\mathbf{H}}, \mathbf{A}) \approx$

$$\frac{1}{L}\sum_{i=1}^{L}\sum_{j=1}^{M}\log_2\left(1 + \frac{|\mathbf{h}_j^{(i)}\mathbf{a}_j|^2}{2\text{Tr}(\mathbf{A}\mathbf{A}^\dagger) + \sum_{l\neq j}|\mathbf{h}_j^{(i)}\mathbf{a}_l|^2}\right).$$

We define $\mathbf{\Delta}^{(i)}$ and $\mathbf{D}^{(i)}$ as in (23) and (24) using the matrix $\mathbf{H}(i)$ instead of $\mathbf{H}$. Using arguments similar to those used in [22], we obtain that the equations $\frac{\partial R}{\partial \mathbf{A}_{ij}} = 0$ imply

$$\sum_{i=1}^{L}\mathbf{H}^{(i)}\mathbf{\Delta}^{(i)} - \mathbf{H}^{(i)\dagger}\mathbf{D}^{(i)}\mathbf{H}^{(i)} - \sigma^2 \text{Tr}(\mathbf{D}^{(i)})\mathbf{A} = 0. \quad (26)$$

Let

$$\mathbf{V} = \sum_{i=1}^{L}\mathbf{H}^{(i)\dagger}\mathbf{D}^{(i)}\mathbf{H}^{(i)} + \sigma^2 \text{Tr}(\mathbf{D}^{(i)})I_M,$$

and

$$T = \sum_{i=1}^{L}\mathbf{H}^{(i)}\mathbf{\Delta}^{(i)}.$$

From (26), we have that

$$\mathbf{A} = \mathbf{V}^{-1}T. \quad (27)$$

This allows us to use the following iterative algorithm for determining $\mathbf{A}$:

1) Assign some initial values to matrices $\mathbf{\Delta}^{(i)}$ and $\mathbf{D}^{(i)}$, for instance $\mathbf{\Delta}^{(i)} = I_M, \mathbf{D}^{(i)} = I_M$
2) Repeat steps 3 and 4 several times
3) Compute $\mathbf{A}$ according to (27);
4) Compute $\mathbf{\Delta}^{(i)}$ and $\mathbf{D}^{(i)}$ according to (23) and (24) using $\mathbf{H}^{(i)}$ instead of $\mathbf{H}$.

*Remark 8:* In numerical simulations (including the ones in Section VI), we have observed that $L = 50$ is sufficient. Further, for typical examples, 4 to 8 iterations are enough. Thus, the number of required iterations is small for numerical convergence in most cases. However, similar to [22], there is no theoretical guarantee on the convergence of the algorithm.

*Remark 9:* The precoding matrix is obtained using numerical techniques. It should be noted that the precoding matrices can be computed offline and implemented using look-up tables. We do not provide the details of this in the paper. Since the precoding is linear, the online computational complexity is low.

### B. Forward Training

The key idea behind sending forward pilots is that users can use these pilots to compute effective channel gains to higher accuracy and reduce the variance of the effective noise. At the same time, we have to spend time for sending forward pilots and a priori it is not clear whether one can obtain any gain from using forward pilots. This motivates us to consider variable number of forward pilots, which can be used for numerical optimization in practical systems. Further, since the users need not estimate the entire channel matrices, we allow for pilot lengths smaller than the number of users.

The base-station transmits $\tau^f$ forward pilots so that every user can obtain estimate of its effective channel gain. Since we are interested in short coherence intervals, we consider the case with very few forward pilots. Note that $\tau^f$ can be less than the number of users $K$. For this reason, we do not restrict to orthogonal pilots in forward training. The forward pilots are obtained by pre-multiplying the vectors $\mathbf{q}_p^{(1)}, \ldots, \mathbf{q}_p^{(\tau^f)}$ with the precoding matrix. In the case of one forward pilot ($\tau^f = 1$), we consider the forward pilots obtained from the vector $\mathbf{q}_p^{(1)} = [1, 1 \ldots]^T$. In the case of $\tau^f = 2$, we consider the forward pilots obtained from the vectors $\mathbf{q}_p^{(1)} = \sqrt{2}[1, 0, 1, 0 \ldots]^T$ and $\mathbf{q}_p^{(1)} = \sqrt{2}[0, 1, 0, 1 \ldots]^T$. It is straightforward to extend this to any number of forward pilots. We denote the vector of forward pilots received by the $k$-th user by $\mathbf{x}_k^p$. The $k$-th user uses $\mathbf{x}_k^p$ to compute $\mathbb{E}[g_{kk}|\mathbf{x}_k^p]$ since variance of $g_{kk} - \mathbb{E}[g_{kk}|\mathbf{x}_k^p]$ contributing to effective noise is smaller that the variance of the corresponding term without forward pilots $g_{kk} - \mathbb{E}[g_{kk}]$.

### C. Achievable Throughput

We use similar techniques (proof is more involved) as in the previous section to obtain net achievable throughput for the transmission method with reverse and forward pilots. From (1), the signal-vector received at the users (all $K$ users) is

$$\mathbf{x}^f = \mathbf{E}^f \mathbf{H}\mathbf{A}\mathbf{q} + \mathbf{z}^f \quad (28)$$

where

$$\mathbf{E}^f = \text{diag}\left\{\left[\sqrt{\rho_1^f} \sqrt{\rho_2^f} \cdots \sqrt{\rho_K^f}\right]^T\right\}.$$

We denote the effective forward channel in (28) by $\mathbf{G} = \mathbf{E}^f \mathbf{H}\mathbf{A}$ with $(i,j)$-th entry $g_{ij}$.

*Theorem 4:* For the transmission method considered, the following downlink weighted-sum rate is achievable during transmission:

$$R_\Sigma = \sum_{k=1}^{K} w_k \mathbb{E}\left[C\left(\frac{|\mathbb{E}[g_{kk}|\mathbf{x}_k^p]|^2}{1 + \sum_{i\neq k}\mathbb{E}[|g_{ki}|^2|\mathbf{x}_k^p] + \mathbf{var}\{g_{kk}|\mathbf{x}_k^p\}}\right)\right] \quad (29)$$

where $C(\theta) = \log_2(1+\theta)$.

*Proof:* In every coherence interval, the $k$-th user receives the vector $\mathbf{x}_k^p$. In the data transmission phase, it receives

$$\begin{aligned}
x_k^f &= g_{kk}q_k + \sum_{i\neq k} g_{ki}q_i + z_k^f \\
&= \mathbb{E}[g_{kk}|\mathbf{x}_k^p]q_k + (g_{kk} - \mathbb{E}[g_{kk}|\mathbf{x}_k^p])q_k \\
&\quad + \sum_{i\neq k} g_{ki}q_i + z_k^f \\
&= \mathbb{E}[g_{kk}|\mathbf{x}_k^p]q_k + \hat{z}_k^f
\end{aligned} \quad (30)$$

where the effective noise

$$\hat{z}_k^f = (g_{kk} - \mathbb{E}[g_{kk}|\mathbf{x}_k^p])q_k + \sum_{i\neq k} g_{ki}q_i + z_k^f.$$

The joint distribution of $\mathbf{x}_k^p$ and $\mathbf{G}$ is known to all users as it depends on the long-term statistics alone (and not the channel realization). In (30), the noise term $\hat{z}_k^f$ is uncorrelated with the signal $q_k$. Note that these terms are not independent, and we do not need independence in the proof. Following the steps used in the proof of Theorem 1, we obtain the achievable rate given in (29). ∎

*Remark 10:* It is (computationally) easy to generate i.i.d. samples from the joint distribution of $\mathbf{x}_k^p$ and $g_{ki}$. Even then computing conditional expectations can be computationally intensive especially for continuous random variables. However, in our setting, we can take advantage of the fact that $z_k^f$ is independent of all other random variables. For example, consider the setting $Y = X + Z$, where $Z$ is an independent random variable with probability density function $f_Z(z)$. In order to compute $\mathbb{E}[X|Y=y]$, we can generate i.i.d. samples of $X$, say $\{x_i\}_{i=1}^L$, and compute

$$\mathbb{E}[X|Y=y] \approx \frac{\sum_{i=1}^L x_i f_Z(y-x_i)}{\sum_{i=1}^L f_Z(y-x_i)}.$$

This idea can be extended to our scenario. Irrespective of this, numerical techniques exist, as it is possible to sample from the joint distribution.

We define net achievable weighted-sum rate as

$$R_{\text{net}} = \max_{\tau^r} \frac{T - \tau^r - \tau^f - 1}{T} R_\Sigma$$

which is consistent with the earlier definition.

In summary, we developed a technique that uses the channel estimate to obtain a precoding matrix that is "good" in expectation for many channel realizations around this estimate. We demonstrate the performance improvement through numerical examples in Section VI.

## V. UPPER BOUND ON SUM RATE

As in the previous sections, we assume that an estimate $\hat{\mathbf{H}}$, the statistics of $\hat{\mathbf{H}}, \tilde{\mathbf{H}}$, and $\mathbf{H}$, and forward SNRs $\rho_k^f$ are available at the base-station. Using this information, the base-station computes a precoding matrix $\mathbf{A}$. The signal received by users is

$$\mathbf{x} = \mathbf{E}^{\mathbf{f}} \mathbf{H} \mathbf{A} \mathbf{q} + \mathbf{z}.$$

As before, we denote the forward pilots received by the $k$-th user using $\mathbf{x}_k^p$. Let

$$C_j = \max_{p(q_j)} I(x_j; q_j | \mathbf{x}_k^p),$$

where $p(q_j)$ is the pdf of $q_j$. The sum capacity is defined by

$$C = C_1 + \ldots + C_K.$$

In Sections III, IV, achievable rates for different communication scenarios were derived. The following simple theorem defines an upper bound on $C$.

*Theorem 5:*

$$C \leq \sum_{j=1}^{K} \log_2\left(1 + \frac{\rho_j^f |\mathbf{h}_j^T \mathbf{a}_j|^2}{1 + \sum_{l\neq j} \rho_l^f |\mathbf{h}_j^T \mathbf{a}_l|^2}\right) \quad (31)$$

*Proof:* Let $\mathbf{G} = \mathbf{H}\mathbf{A}$. Then,

$$\begin{aligned}
C_j &= \max_{p(q_j)} I(x_j; q_j | \mathbf{x}_k^p) \\
&\leq \max_{p(q_j)} I(x_j \mathbf{G}; q_j | \mathbf{x}_k^p) \\
&= \max_{p(q_j)} \{I(x_j; q_j | \mathbf{G}, \mathbf{x}_k^p) + I(\mathbf{G}; q_j | \mathbf{x}_k^p)\} \\
&= \max_{p(q_j)} I(x_j; q_j | \mathbf{G}) \\
&= \log_2\left(1 + \frac{\rho_j^f |\mathbf{h}_j^T \mathbf{a}_j|^2}{1 + \sum_{t\neq j} \rho_t^f |\mathbf{h}_j^T \mathbf{a}_t|^2}\right).
\end{aligned}$$

Here, we used the facts that $\mathbf{G}$ and $q_j$ are independent and therefore $I(\mathbf{G}; q_j | \mathbf{x}_k^p) = 0$, and that $\mathbf{x}_k^p$ is a noisy version of $\mathbf{G}$ and therefore $I(x_j; q_j | \mathbf{G}, \mathbf{x}_k^p) = I(x_j; q_j | \mathbf{G})$. ∎

It is easy to see that the same bound is valid if no forward pilots are available to users. In general this upper bound is valid for any particular scheme of generating precoding matrix $\mathbf{A}$. Hence, the bound can be used in all communications scenarios considered in the previous sections. In this way, we can obtain an upper bound on the sum rate of any specific communication scenario and any specific precoding method. In the numerical results presented in the next section, we demonstrate that the gap between our achievable rates derived in the previous sections, and the corresponding upper bound is quite narrow.

Instead of using a specific precoding method in Theorem 5, we can try to use a precoding matrix $\mathbf{A}$ that maximizes (31), under assumption that only $\hat{\mathbf{H}}$, the statistics of $\hat{\mathbf{H}}, \tilde{\mathbf{H}}$, and $\mathbf{H}$, and forward SNRs $\rho_k^f$ are available at the base-station. This would give us an upper bound that is not dependent on a specific precoding method. In the case that such an upper bound is close to the achievable sum rate of some specific precoding method, we could claim that we have not only closely identified the sum rate of that specific precoding method, but also that the scheme itself is close to optimal linear precoding.

The problem of finding a precoding matrix $\mathbf{A}$ that provably maximizes (31), especially in the case when the true channel matrix $\mathbf{H}$ is not available, looks to be very hard. We suggest the following approximate approach. The algorithm described in Section IV-A allows us to find, approximately, $\mathbf{A}$ that provides a local maximum for $\mathbb{E}_{\tilde{\mathbf{H}}}[R(\hat{\mathbf{H}} + \tilde{\mathbf{H}}, \mathbf{A})]$. Running the algorithm several times, with distinct random matrices for $\Delta$ and $D$ in step 1, we can find several, say a hundred, local maxima of $\mathbb{E}_{\tilde{H}}[R(\hat{H} + \tilde{H}, A)]$. Let C-UB-Opt be the maximum of these local maxima. Though, strictly speaking, C-UB-Opt is not the global maximum of $\mathbb{E}_{\tilde{H}}[R(\hat{H} + \tilde{H}, A)]$, it is likely that there is no linear precoding method that would significantly outperform C-UB-Opt. In the next section, we will use C-UB-Opt as a scheme independent upper bound for some communication scenarios.

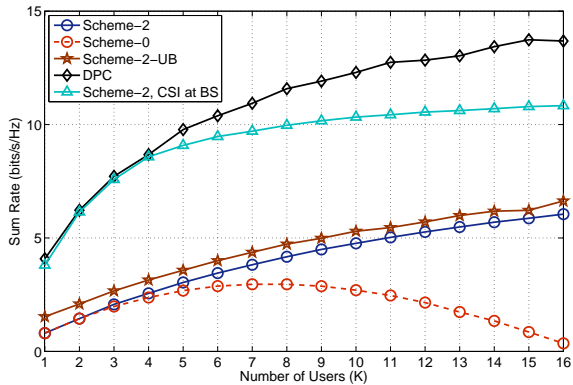

Fig. 4. Achievable sum rate for forward SNR of 0 dB and reverse SNR of $-10$ dB

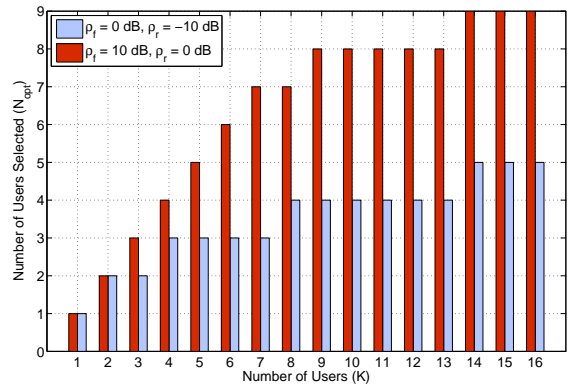

Fig. 5. Number of selected users versus total number of users

## VI. NUMERICAL RESULTS

Scheme-UB refers to the upper bound obtained by assuming perfect knowledge of the effective channel matrix at the users. Note that this is a scheme dependent upper bound. We have conducted extensive simulations for various system parameters, and the observations provided are based on these simulations. However, we provide only few representative numerical results here.

### A. Training on Reverse Link Only

We consider this transmission method in the communication regime when SNRs are low. Scheme-0 denotes ZF precoding method and Scheme-1 denotes the generalized ZF precoding method with optimized $p_i$ values but no user selection. Scheme-2 denotes the method where user selection is used along with Scheme-1. Scheme-1 and Scheme-2 are techniques developed in this paper. Scheme-0 refers to the scheme in [21].

*1) Homogeneous Users:* For homogenous users, Scheme-1 is identical to Scheme-0. First, we keep the training sequence length equal to the number of users, i.e., $\tau^r = K$. This setting clearly is the minimum channel training overhead. In Figure 4, we plot sum rate versus the number of users $K = \{1, 2, \ldots, M\}$ for $M = 16$ when forward SNR $\rho^f = 0$ dB and reverse SNR $\rho^r = -10$ dB. In addition to Scheme-0 and Scheme-2 sum rates, we plot upper bound obtained according to Theorem 5, Scheme-2 performance when CSI is available at the base-station, and the DPC upper bound. The reduction in sum rate due to lack of full CSI at base-station is significant. As expected, the performance of DPC is significantly better compared to linear precoder especially when

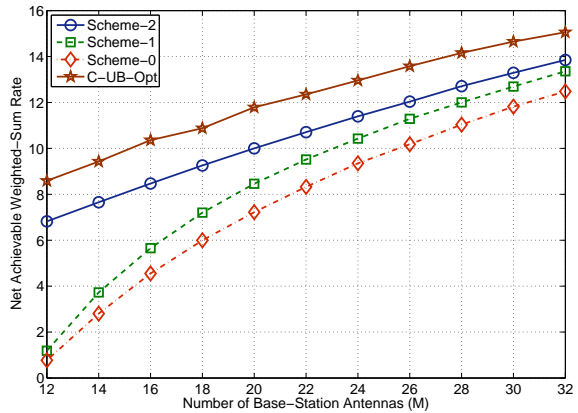

Fig. 6. Net achievable weighted-sum rate for a system with 12 users

$M = K$. Now onwards, we do not compare with DPC as our focus is on linear precoders with channel imperfections. Since the gap between the Scheme-2 sum rate and Scheme-2 upper bound is relatively small, the restriction to training on reverse link only is not significant for the SNRs considered here. We observe that the user selection strategy used in Scheme-2 gives significant improvement over existing Scheme-0. In Figure 5, we plot the number of users selected by Scheme-2 $N_{opt}$ versus the number of users present $K$ for different SNRs (mentioned in the plot) and $M = 16$.

*2) Heterogeneous Users:* We consider coherence interval $T = 30$ symbols and 12 users with forward SNRs $\{0, 0, 0, 5, 5, 5, 5, 5, 5, 10, 10, 10\}$ dB. The reverse SNR associated with every user is considered to be 10 dB lower than its forward SNR. All users are assigned unit weights. We plot the net achievable sum rate versus $M$ for this system in Figure 6. The improvement obtained using modified ZF precoding with optimized $p_i$ values is significant. We remark that the performance gain due to user selection is very significant when the number of users are comparable to the number of base-station antennas.

*3) Optimal Training Length:* We consider a homogeneous system with $M = 32$ antennas at the base-station, $K = 8$ users and coherence interval of $T = 30$ symbols. For Scheme-

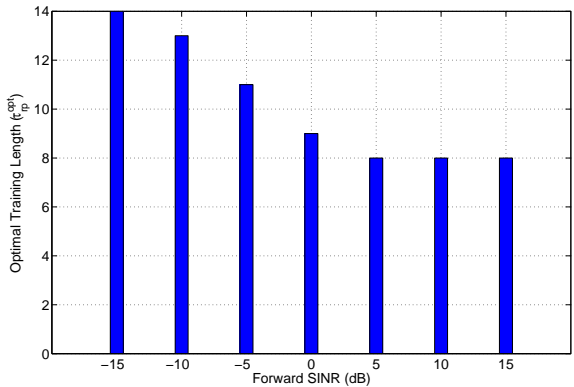

Fig. 7. Optimal training length versus forward SNR

TABLE I
COMPARISON OF VARIOUS SCHEMES

| $\rho^f$ (dB) | 5 | 10 | 15 | 20 | 25 | 30 |
|---|---|---|---|---|---|---|
| ZF-FP(0) | 0.65 | 1.93 | 4.95 | 8.54 | 12.12 | 13.68 |
| ZF-UB | 1.22 | 2.89 | 6.42 | 11.97 | 19.10 | 27.62 |
| ZF-Sch-FP(0) | 3.87 | 7.32 | 11.37 | 15.06 | 17.88 | 19.08 |
| ZF-Sch-FP(1) | 2.59 | 5.38 | 9.39 | 13.27 | 19.64 | 26.22 |
| ZF-Sch-FP(2) | 3.50 | 6.64 | 10.21 | 15.09 | 20.19 | 26.69 |
| ZF-Sch-UB | 4.74 | 8.42 | 13.39 | 19.33 | 25.83 | 32.71 |
| SVH-FP(1) | 3.27 | 6.38 | 10.74 | 15.69 | 21.87 | 27.16 |
| SVH-FP(2) | 3.71 | 6.95 | 10.98 | 16.17 | 21.33 | 27.15 |
| SVH-UB | 5.30 | 9.54 | 14.78 | 20.97 | 27.49 | 34.07 |
| Mod-SVH-FP(1) | 3.33 | 6.54 | 10.62 | 16.92 | 22.44 | 29.45 |
| Mod-SVH-FP(2) | 3.51 | 7.27 | 11.22 | 15.42 | 20.54 | 26.67 |
| Mod-SVH-UB | 5.34 | 9.71 | 15.28 | 21.57 | 28.25 | 35.06 |

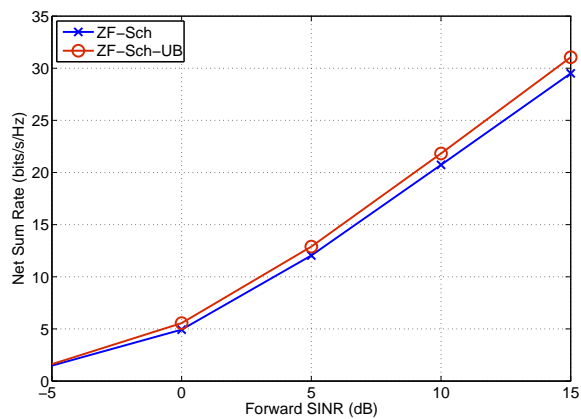

Fig. 8. Net sum rate versus forward SNR

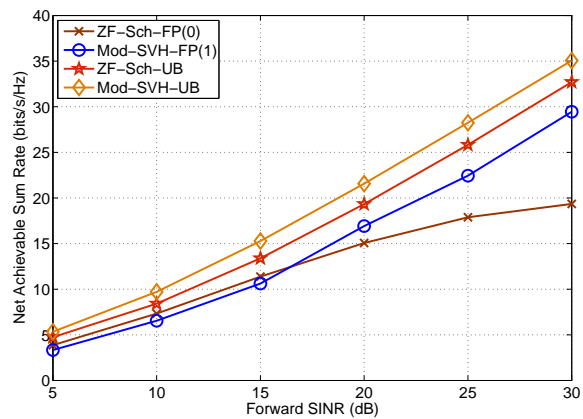

Fig. 9. Net rate versus forward SNR for $M = K = 8$

2, we obtain the optimal training length and the net sum rate for different values of forward SNR through brute-force optimization. For every forward SNR considered, we take the reverse SNR to be 10 dB lower than the corresponding forward SNR. We plot the optimal training lengths in Figure 7 and net sum rates in Figure 8. The behavior of optimal training length with reverse SNR is as predicted in Section III-F - $T/2$ in low SNR regime and $K$ in high SNR regime. In Figure 8, we denote ZF with user selection (scheduling) by ZF-Sch and the corresponding upper bound by ZF-Sch-UB.

### B. Training on Reverse and Forward Links

We consider this transmission method for moderate to high SNRs. We use FP($n$) to denote a precoding method using $n$ number of forward pilots. Note that FP(0) denotes training on reverse link only. We denote results obtained with zero-forcing by ZF, zero-forcing with user selection by ZF-Sch, the approach in [22] by SVH and the modified algorithm given in Section IV-A by Mod-SVH. We compare the performance of different methods using numerical examples. For the algorithm Mod-SVH, we use the value $L = 50$ in the simulations and 5 iterations. We observe that these 5 iterations is enough to provide numerical convergence (i.e., a reasonable error bound) in our examples.

We consider a system with $K = 8$ users, $M = 8$ antennas at the base-station, reverse training length of $\tau^r = 8$ and coherence interval of $T = 30$ symbols. We consider the following example. We keep the value of reverse SNR 10 dB lower than the forward SNR. For the different methods considered, we obtain the achievable sum rate for forward SNRs ranging from 5 dB to 30 dB. These sum rates are given in Table VI-B. We plot the methods ZF-Sch-FP(0) and Mod-SVG-FP(1) in Figure 9. We observe significant improvement in net rate by utilizing forward pilots at high forward SNRs. In addition, it is interesting to note that we perform reasonably close to the upper bound by using one or two forward pilots.

### VII. CONCLUSION

We develop a general framework to study downlink TDD systems that account for channel training overhead and channel estimation error. In contrast to the limited-feedback framework for FDD systems, we account for all channel training overhead in the overall system throughput. In the first part of the paper, we focus on downlink systems with large number of antennas at the base-station. We clearly demonstrate the advantage of TDD operation in this setting. In particular, with increasing number of base-station antennas, the TDD operation helps in improving the effective forward channel without affecting the training sequence length required. We present a generalized zero-forcing precoding method in this setting. We use a combination of convex optimization based technique and opportunistic user selection to maximize the overall system throughput. In the second part of the paper, we consider the general setting, i.e., we do not limit focus to downlink systems with large number of base-station antennas. We present a linear precoding method than results from an approach to find a local maximum for a non-convex optimization problem that is related to the system throughput. Through simulations, we show that these precoding schemes provide good overall system throughput.

ACKNOWLEDGMENT

The authors would like to thank T. L. Marzetta for helpful discussions on this topic.